\documentclass[epj,referee]{svjour}

\usepackage{mmaptm}
\usepackage{times}

\usepackage{amsmath}
\usepackage{amssymb}

\usepackage{graphicx}
\usepackage{psfrag}

\setkeys{Gin}{keepaspectratio=true}

\newcommand{\xlabel}[1]{\psfrag{xlabel}[t][]{#1}}
\newcommand{\ylabel}[1]{\psfrag{ylabel}[B][]{#1}}

\DeclareMathAlphabet{\mathrmb}{OT1}{cmr}{b}{n}
\DeclareMathAlphabet{\mathsfb}{OT1}{cmss}{bx}{n}\renewcommand\ensuremath\relax
\newcommand\textfrac[2]{\frac{\textstyle #1}{\textstyle #2}}
\newcommand\Frac[2]{\frac{\displaystyle #1}{\displaystyle #2}}
\newcommand{\Vector}[1]{\ensuremath{\mathrmb{#1}}}

\newcommand{\der}[2]{\Frac{d#1}{d#2}}

\newcommand{\Int}[4]{\int_{#1}^{#2}\!\!{#3}d{#4}}

\newcommand{\OrderOf}[1]{\ensuremath{O(#1)}}

\newcommand{\etal}{\emph{et al.}}
\begin{document}

\title{Analysis of radial segregation of granular mixtures in a
 rotating drum}
\author{Saikat Chakraborty\inst{1} \and Prabhu R. Nott\inst{1}
\thanks{E-mail: prnott@chemeng.iisc.ernet.in} \and J. Ravi Prakash\inst{2}}
\institute{Department of Chemical Engineering, Indian Institute of Science,
Bangalore 560$\,$012, INDIA \and Department of Chemical Engineering,
Indian Institute of Technology Madras, Chennai 400$\,$036, INDIA}
\date{Received: date / Revised version: date}
\abstract{
This paper considers the segregation of a granular mixture in a rotating
drum. Extending a recent kinematic model for grain transport on sandpile
surfaces to the case of rotating drums, an analysis is presented for
radial segregation in the rolling regime, where a thin layer is
avalanching down while the rest of the material follows rigid body
rotation. We argue that segregation is driven not just by differences in
the angle of repose of the species, as has been assumed in earlier
investigations, but also by differences in the size and surface properties
of the grains.  The cases of grains differing only in size (slightly or
widely) and only in surface properties are considered, and the predictions
are in qualitative agreement with observations.  The model yields results
inconsistent with the assumptions for more general cases, and we speculate
on how this may be corrected.
} 
\maketitle

\section{Introduction}\label{intro}
        The production of many goods, ranging from pharmaceuticals and
foods to polymers and semiconductors, depends on reliable uniform mixing
of granular materials. Although there have been several recent advances,
particulate mixing is poorly understood, and one cannot \emph{a priori}
predict the effectiveness of any mixing process. Indeed, mixing
operations often result in segregation or de-mixing, and even the
parameters that control mixing and segregation are not fully understood.

        Rotating drums, or kilns, are employed in industry to carry out
a range of operations; some examples are the calcination of limestone,
reduction of oxide ore, clinkering of cementitious materials, waste
incineration and calcination of petroleum coke. Owing to its industrial
importance the rotary kiln has been the subject of numerous investigations.
Significant improvement in kiln performance may be achieved by better
understanding grain transport during operation.

        In a  rotating drum  several regimes of  bed motion  have been
identified, namely, slipping, slumping, rolling, cascading, cataracting
and centrifuging \cite{henien_etal}. The most desirable bed motion for
many industrial operations is the rolling mode, as it helps in promoting
good mixing of the particles along with rapid renewal of the exposed
material. The rolling regime is characterized by two distinct regions:
the ``active'' and ``passive'' layers \cite{boateng_barr96}. Grains in
the passive layer execute rigid body rotation along with the drum and
the streamlines are circular (see figure 1). The active layer is usually
very thin in comparison with the extent of the passive layer. Grains
roll down rapidly in the active layer and streamlines have been observed
to be straight \cite{nityanand_etal}.
\begin{figure}
\includegraphics[width=8cm]{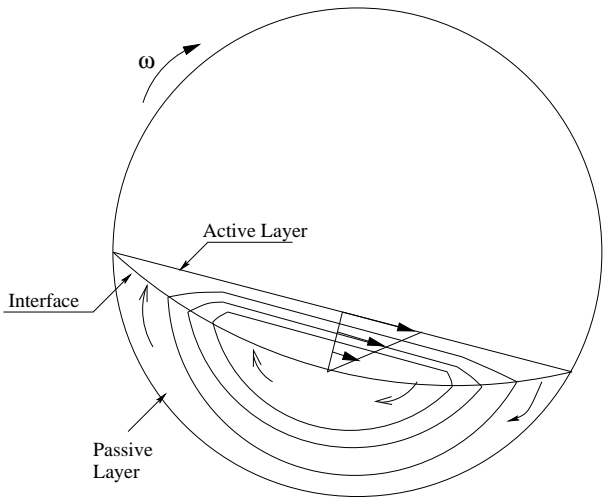}
\vspace{0cm}
\caption{Streamlines in the active and passive layers in the rolling
regime of bed motion in a rotating drum.  Grains cascade rapidly down
the active layer, which is usually very thin in comparison with the
passive layer.}
\label{stream}
\end{figure}

        While rotating drums are used to  mix particles of varying sizes
and shapes to obtain a homogeneous mixture necessary for certain industrial
processes, there is considerable evidence of segregation when the charge is
a granular mixture of different properties. A mixture of grains differing
either in size and/or roughness \cite{metcalfe_etal}-\cite{clement_etal},
or density \cite{cantelaube_bideau} when rotated in a drum is seen to
undergo radial segregation, characterized by the formation of a core of
smaller or rougher grains. This radially segregated core is seen to evolve
into alternate bands of larger (or smoother) and smaller (or rougher)
grains along the length of the drum \cite{dasgupta_etal,nakagawa_etal}.
Axial segregation is found to occur over a considerably larger time scale
than radial segregation, which is usually complete within a few drum
rotations. In a rotary kiln, radial segregation could lead to poor contact
between the gas flowing above and the particles in the core, which would
result in poor heat transfer and/or lower rate of reaction. Axial
segregation, would lead to products of fluctuating qualities.

        Analysis of  granular segregation has been  undertaken in some
recent studies \emph{via} discrete element simulations \cite{khakhar_etal},
and using coarse-grained continuum models
\cite{khakhar_etal}-\cite{boateng_barr97}. A recent approach, which is the
one followed in this work, is based on the theoretical formalism of
Bouchaud, Cates, Ravi Prakash and Edwards \cite{bcre}. Henceforth referred
to as BCRE in this work, this study models grain transport on a sandpile
surface, and has been employed in \cite{boutreux_degennes} for studying
segregation during the filling of a silo with a mixture of two species. The
``minimal'' model of \cite{boutreux_degennes} describes the case of grains
with equal sizes but with small differences in surface properties. The same
formalism was used in \cite{makse97} to demonstrate the possibility of
complete segregation in a mixture of large smooth grains and small rough
grains, and spontaneous stratification, \emph{i.e.} alternating layers of
the two species, for a mixture of large rough grains and small smooth
grains. The minimal model of \cite{boutreux_degennes} was generalised by
\cite{boutreux_etal} to accommodate grains differing slightly in surface
properties as well as size.

        In this work, we  extend the formalism of \cite{boutreux_etal}
to address the problem of segregation in rotating drums of a granular
mixture whose constituents differ in size and surface properties. The
problem of segregation in rotating drums has been addressed in
\cite{makse98} for some special cases; however, our approach differs
from theirs in some fundamental ways, which we will point out in this
paper.

\section{ The model}
\label{model}

        Following BCRE,  we assume the existence of  a sharp interface
between the active and passive layers. Grains in the active layer are
referred to as rolling grains and those in the passive layer as
immobile. Collisions between rolling and immobile grains occur at the
interface, which lead to exchange of grains between the two regions. We
are interested in studying the case of a bi-disperse mixture consisting
of $\alpha$ and $\beta$ type grains, which differ in surface properties
and/or size.

        We call $R_{\alpha}(x,t)$ and $R_{\beta}(x,t)$ the number
of $\alpha$ and $\beta$ type rolling grains in the active layer per unit
length of the interface, and $R(x,t) \equiv R_{\alpha}(x,t) +
R_{\beta}(x,t)$ the total number of rolling grains (per unit length) in
the active layer. The local slope of the interface is $\theta(x,t)$, the
thickness of the active layer is $\delta(x,t)$ and the height of the
passive layer is $h(x,t)$ (see figure 2). As the material in the passive
layer only undergoes solid body rotation, it is sufficient to specify
the the concentration at the top surface to fully specify the
concentration throughout this layer. Consequently, we define the number
fractions of immobile grains in the passive layer
\emph{at the interface}\/ $\phi_{\alpha}(x,t)$ and $\phi_{\beta}(x,t)$.
The surface of the active layer is inclined at an angle $\theta_r$, the
dynamic angle of repose), with the horizontal.

        The conservation  equations for the two species  in the active
layer are
\begin{eqnarray}
\partial_{t}{R}_{\alpha}(x,t) & = &
-\partial_{x}(v(x,t)R_{\alpha}(x,t))  + \Gamma_{\alpha} \nonumber \\
 & & + \partial_{x} (  D \partial_{x} R_{\alpha}(x,t)) \label{e3} \\
\partial_{t}{R}_{\beta}(x,t) & = &
-\partial_{x}(v(x,t)R_{\beta}(x,t)) + \Gamma_{\beta} \nonumber \\
 & & +  \partial_{x} ( D \partial_{x} R_{\beta}(x,t))
\end{eqnarray}
where $v(x,t)$ and $D$ are the mean velocity and the diffusivity of the
grains, respectively. The active layer is assumed to be very thin, and
variations across it are not resolved in this model. All the interesting
physics in this model is buried in $\Gamma_{\alpha}$ and $\Gamma_{\beta}$,
which are the rate of conversion of immobile grains of each type into
rolling grains by collisions at the interface, and we will specify their
functional form shortly.

        In the passive layer, grain transport is only by advection due
to the rotation of the drum. The equation for conservation of $\alpha$ type
grains, integrated across the passive layer, is
\begin{equation}
\label{e4}
\rho_{n} \, \partial_{t}\,  \Int{0}{h} {\phi^{p}_{\alpha}(x,y,t)}{y} =
\phi_{\alpha} \rho_{n} \Vector{v_{\theta}}.\Vector{n} - \Gamma_{\alpha},
\end{equation}
and similarly for $\beta$. Here, $\phi^{p}_{\alpha}(x,y,t)$ is the
number fraction of $\alpha$ type grains \emph{inside} the passive layer,
$\Vector{v_{\theta}}$ is the velocity field due to solid body rotation,
$\rho_{n}$ is the number density of grains in the passive layer (taken
to be constant), and $\Vector{n}$ is the unit normal to the interface,
pointing from the passive layer to the active. (Due to the circular
streamlines in the passive layer $\phi^{p}_{\alpha}$ can be related to
the number fraction at the interface $\phi_{\alpha}$.) The conservation
for all grains then assumes the form
\begin{equation}
\label{e5}
\rho_{n} \, \partial_{t} h \; = \;
\rho_{n} \Vector{v_{\theta}}.\Vector{n} - \sum_{i=\alpha, \beta}
\Gamma_{i}
\end{equation}
\begin{figure}
\includegraphics[width=8cm]{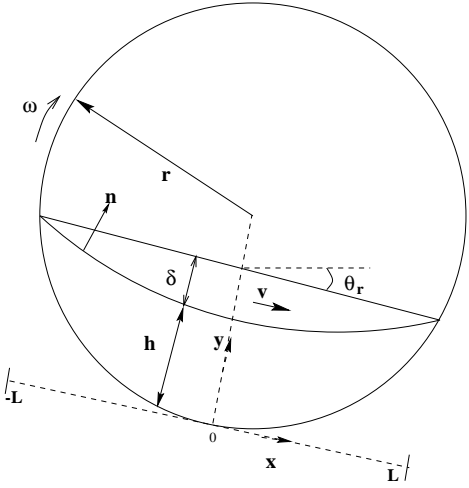}
\vspace{0cm}
\caption{The coordinate frame for our model equations,
showing some of the variables of interest}
\label{symbol}
\end{figure}

\subsection{Model for $\Gamma_i$} \label{formu:canon}

        As in  BCRE, we write  an expression for  $\Gamma_{\alpha}$ in
terms of the rolling grain densities $R_i$ and the number fraction of
immobile grains $\phi_i$ at the interface. Boutreux and de Gennes (1996)
consider the four possible outcomes of binary collisions between rolling
and immobile grains at the interface; \emph{amplification, capture,
exchange and recoil}. Amplification of species $\alpha$ refers to the event
of an $\alpha$ type immobile grain being dislodged into the active layer by
a rolling grain. This could happen either by
\emph{auto-amplification}, when the colliding grain is also of type
$\alpha$, or by \emph{cross-amplification}, when it is of type $\beta$. The
auto-amplification (cross-amplification) rate of $\alpha$ type species will
increase with $R_{\alpha}$ ($R_{\beta}$), $\phi_{\alpha}$ and the collision
frequency $f$. Capture of species $\alpha$ refers to the event of an
$\alpha$ type rolling grain being rendered immobile and absorbed by the
passive layer. Arguing similarly for the rate of capture (and noting that
exchange and recoil leave the net transfer of grains unchanged), we write
the simplest form for $\Gamma_{\alpha}$ as
\begin{eqnarray}
\Gamma_{\alpha} & = & \gamma_{\alpha \alpha}^{a}
\Frac{v}{d_{p}} R_{\alpha}  \phi_{\alpha} g^{a}(\psi) + \gamma_{\alpha
        \beta}^{a} \Frac{v}{d_{p}} R_{\beta} \phi_{\alpha}
        g^{a}(\psi) \nonumber \\
        & & - \gamma_{\alpha}^{c} \Frac{v}{d_{p}} R_{\alpha} g^{c}(\psi)
\label{e9}
\end{eqnarray}
wherein we have taken the collision frequency, $f$, to vary as
$\textfrac{v}{d_p}$, $d_p$ being the average grain diameter. The
coefficients $\gamma_{\alpha \alpha}^{a}$, $\gamma_{\alpha \beta}^{a}$ and
$\gamma_{\alpha}^{c}$ are constants of order unity and $g^a(\psi)$ and
$g^c(\psi)$ functions of the slope of the interface.

We  note  that  unlike  in  previous  studies  investigating  granular
segregation \cite{boutreux_degennes,boutreux_etal,makse98},  we do not
set the  velocity $v$ to  a constant, but  allow it to vary  along the
length  of  the  active  layer  by solving  the  momentum  balance  in
conjunction with the species balances (\ref{e3}).

        To   complete  the   formulation,  the   functional   forms  of
$g^a(\psi)$ and $g^c(\psi)$ must be specified. Recognizing that the rates
of amplification and capture depend not just on the local slope of the
interface but on the \emph{difference} between the local slope and the
angle of repose, we set $\psi = \theta (x) - \theta_{r} (x)$, where
$\theta(x)$ is local slope of the interface and $\theta_{r} (x)$ is the
angle of repose for the mixture. Next, $\Gamma_{\alpha}^{a}$
($\Gamma_{\alpha}^{c}$) should rise (diminish) with $\psi$, but we do not
expect it to vanish altogether at $\psi = 0$, since we expect a fraction of
the grains striking a surface at the angle of repose to dislodge other
grains (get captured). However, for large negative (positive) values of
$\psi$, the rate of amplification (capture) should become negligibly small.
In accordance with these arguments, we assume the following continuous
forms:
\begin{eqnarray}
\label{e6}
        g^{a}(\psi) & = & \Frac{A  \: \exp(B \psi)}{1 + \exp(B \psi)}\\
        g^{c}(\psi) & = & \Frac{A}{1 + \exp(B \psi)}
\end{eqnarray}
where $A$ and $B$ are constants of \OrderOf{$\textfrac{d_{p}}{L}$} and
\OrderOf{$\textfrac{L}{d_{p}}$}, respectively. The notable feature of the above
forms is that $g^a + g^c$ is constant - while we are unable to ascribe a
physical meaning to this constraint now, it is essential if the solutions
are to be symmetric about $x=0$. The issue of symmetry is commented on in
greater detail in section \ref{symmetry}.

        We note here that Makse\cite{makse98} considered special cases
of the above slope functions for very small and very large differences
of angles of repose. He also assumed that differences in size and
surface properties between the species result only in differences in
their angles of repose, which is the ultimate cause of segregation, and
that the coefficients for auto-amplification and cross-amplification
($\gamma_{\alpha \alpha}^{a}$ and $\gamma_{\alpha \beta}^{a}$) are
equal. We believe that ascribing segregation purely to a difference in
the angle of repose is unrealistic; it surely does not address
segregation of grains of different sizes, but made of the same material.
In this case, segregation arises because the probability of a larger
grain dislodging a smaller one is greater than that of the larger one
dislodging another larger one, as we shall demonstrate below.

        We propose that segregation due to size and surface properties are
driven primarily by amplification and capture respectively, as
elaborated in section \ref{pred:surf}. This has been done by allowing
(unlike \cite{makse98}) the coefficients $\gamma^{a}_{\alpha \alpha}$,
$\gamma^{a}_{\alpha \beta}$, $\gamma^{c}_{\alpha}$ etc. to depend on the
size, roughness and type of interacting grains. In addition, the effect
of differences of angle of repose caused by differences on size or
surface properties has been incorporated through the slope functions
(eq. \ref{e6}). Our results, shown in sections \ref{pred:surf} --
\ref{pred:perc}, illustrate quite clearly that the qualitative nature of
size and roughness segregation is different, which is in agreement with
previous experimental observations \cite{henien_etal} --
\cite{cantelaube_bideau}. We allow the the angle of repose to depend on
the grain concentration in our model, but assume it to be constant in
this paper for the sake of simplicity.

\subsection{Momentum Balance} \label{form:mom}

        Previous studies that used the same theoretical framework made
the unrealistic assumption that the velocity of grains in the active is
constant -- clearly, the velocity must start with zero at the top end of
the active layer, reach a maximum at an intermediate point, and again
vanish at the lower end of the active layer. We allow the velocity to vary
along the length of the layer, and write a momentum balance for the rolling
grains, averaged over the thickness of the active layer:
\begin{eqnarray}
\label{mom1}
\partial_{t}\:(R\,v)\;+\;\partial_{x}\:(R\,v^{2}) & = & v\:\sum_{i=\alpha,\beta}
\Gamma_{i}\;+\;R\:g\:\sin\theta_r \nonumber \\
& & - \sigma_f - \sigma_c.
\end{eqnarray}
The first term in the right hand side accounts for $x$ momentum influx due
to conversion of immobile grains into rolling grains, the second is the
momentum generated due to the gravitational body force, and the third and
fourth terms are the shear stresses at the interface due to frictional
grain interactions at the interface and grain collisions
\cite{johnson_etal}, respectively. We assume a Coulombic form for the
frictional stress,
\begin{equation*}
\sigma_{f}\;=\; \mu \:R\:g\:\cos\theta_r\; ,
\end{equation*}
where $\mu$ is the coefficient of friction of the material. The collisional
shear stress, $\sigma_{c}$, is generated by collisions of the exchange and
recoil type, discussed earlier, that do not lead to inter-conversion of
grains but do result in loss of momentum. Therefore, we choose a
constitutive form of $\sigma_{c}$ that is similar to the momentum flux due
to inter-conversion,
\begin{equation*}
\sigma_{c}\;=\;\lambda \:R \: \Frac{v^{2}}{d_{p}},
\end{equation*}
where $\lambda$, like $A$, is a constant of proportionality of
\OrderOf{$\textfrac{d_{p}}{L}$}.

\subsection{Scaling and Leading Order Analysis} \label{formu:scale}

        To clearly illustrate the  relative magnitude of  the different
terms in  our model equations,  we define the  following dimensionless
variables,
\begin{equation*}
x^{\ast}\:=\:\Frac{x}{L},\:\;v^{\ast}\:=\:\Frac{v}{\omega\,L},\:\;
\delta^{\ast}\:=\:\Frac{\delta}{d_{p}},\:\;R_{\alpha}^{\ast}\:=\:
\Frac{R_{\alpha}}{\rho_{n}\,L},\:\:h^{\ast}\:=\;\Frac{h}{L}
\end{equation*}
and the following dimensionless parameters,
\begin{equation*}
\epsilon\:=\:\Frac{d_p}{L},\:\;r^{\ast}\:=\:\Frac{r}{L},\:\;D^{\ast}\:=\:\Frac{D}{\omega\,L\,d_{p}},
\:\;g^{\ast}\:=\:\Frac{g}{\omega^{2}\,L},
\:\; \lambda^{\ast}\:=\: \Frac{L \,\lambda}{d_{p}}
\end{equation*}
\noindent where  $L$ is the half-length of  the free-surface, $\omega$
is the rotational  velocity of the drum, $d_{p}$  is the average grain
size, $r$  is the radius of the  drum and $\rho_{n}$ is  the number of
grains per unit area of the passive layer.

        The  active layer is expected to be only a few grains deep,
and its thickness $\delta$ is therefore scaled with $d_{p}$. The
diffusion coefficient, $D$, being a material property scales as the
product of the velocity and the grain size. Using the above normalized
variables, and dropping the asterisks for the sake of simplicity,
(\ref{e3}) tranforms to the following dimensionless form at steady
state:
\begin{equation}
\label{e15}
\der{(v \, R_{\alpha})}{x}\;=\;\epsilon\:D\:\der{^{2}
  R}{x^2}\;+\;\Gamma_{\alpha}
\end{equation}
with,
\begin{eqnarray}
\label{e16}
 \Gamma_{\alpha} & = & \gamma_{\alpha \alpha} \:v
  \:R_{\alpha} \:\phi_{\alpha}
  \:f^{a}(\psi) \; +
  \;\gamma_{\alpha \beta} \;v  \:R_{\beta} \:\phi_{\alpha}
  \:f^{a}(\psi)\; \nonumber \\
& & - \; \gamma_{\alpha}^{c} \:v \:R_{\alpha}\:f^{c}(\psi)
\end{eqnarray}
\begin{equation}
f^{a}(\psi)\;=\; \Frac{\exp(\psi)}{1\;+\;\exp(\psi)}
\; ;
f^{c}(\psi)\;=\; \Frac{1}{1\;+\; \exp(\psi)}\; ,
\label{e17}
\end{equation}
where
\begin{equation*}
\psi \;=\; \der{\delta}{x}
\end{equation*}
\noindent The above forms of the slope functions result upon setting
$A L/d_p\, =\, B d_p/L\, =\, 1$, which we have done for simplicity.

        We then consider the case of small $\epsilon$, as is the case
in most practical situations, and seek the to leading order solution. In
this approximation, the governing equations reduce to the following set:
\begin{equation}
\label{final1}
\der{(v\,R)}{x}\;=\; \sum_{i=\alpha, \beta}\Gamma_{i} \; ,
\end{equation}
\begin{equation}
\label{final2}
\der{(v \, R_{\alpha})}{x}\;=\;\Gamma_{\alpha} \; ,
\end{equation}
\begin{equation}
\label{final3}
- \, x \, = \, \sum_{i=\alpha, \beta} \Gamma_{i} \; ,
\end{equation}
\begin{equation}
\label{final4}
- \, \phi_{\alpha} \, x \, = \, \Gamma_{\alpha} \; ,
\end{equation}
\begin{eqnarray}
\label{final5}
\der{(R \, v^{2})}{x}& = & v\:\sum_{i=\alpha, \beta}
\Gamma_{i}\;+\;R\:g\:[\,\sin\theta_r\;-\;\mu\:\cos\theta_r\,] \nonumber \\
& & - \;\lambda \: R \: v^{2} \; ,
\end{eqnarray}
with
\begin{equation}
\label{final6}
R_{\alpha}\;+\;R_{\beta}\; \equiv \;R\, ,
\end{equation}
\begin{equation}
\label{final7}
\phi_{\alpha}\;+\; \phi_{\beta}\; \equiv 1 .
\end{equation}

        An additional condition is  required for solving for
$\phi_{\alpha}$, and this is the loading condition specifying the total
amount of material loaded into the drum,
\begin{equation*}
  \Int{-L}{L}{\Int{y_d}{y_a}{\phi_{\alpha}^p}{y}}{x} \: = \:
V\,\Phi_{\alpha},
\end{equation*}
where V and $\Phi_{\alpha}$ are the total volume and number fraction of
$\alpha$ type grains in the entire charge, respectively. Since
$\phi_{\alpha}^p$ is constant along each circular streamline in the
passive layer, the above condition reduces at leading order to the
following dimensionless form:
\begin{equation}
\label{load}
  \Int{-1}{1}{\phi_{\alpha}(x) \, x}{x} \: = \: \Phi_{\alpha}.
\end{equation}

\section{Model Predictions \label{c:pred}}

        Substituting  (\ref{final3}) in  (\ref{final1}),  and applying
the boundary condition $v\,R\:=\:0$ at $ x \: = \: -1 $, we get
\begin{equation}
\label{pred2}
  v \, R \: = \: \Frac{(1\,-\,x^{2})}{2}.
\end{equation}
Substituting (\ref{pred2}) in (\ref{final5}), and solving for the
velocity subject to the boundary condition $v \: = \: 0 $ at $x \: =
\: -1$, we get
\begin{equation}
\label{pred4}
  v \; = \; \sqrt{\Frac{C}{\lambda} \: ( \, 1 \; - \; \exp[- \lambda \:
    (\, 1 \: + \: x)])},
\end{equation}
where
\begin{equation*}
  C \; = \; g \, [ \, \sin\theta_r \; - \; \mu \, \cos\theta_r \,].
\end{equation*}
The above velocity profile does not vanish at the lower end of the active
layer, $y=1$; this boundary condition may be enforced by putting in a bulk
viscosity for the flowing medium. The correction to the velocity would be
of higher order in $\epsilon$ everywhere, except in a thin boundary layer
near $y=1$. Also notable is that while the velocity is not symmetric about
$x=0$, the flux $v\, R$ is.

        Substituting (\ref{pred2}) and (\ref{final4}) in (\ref{final2})
and simplifying yields
\begin{equation}
\label{pred7}
  \der{u_{\alpha}}{x} \: = \: \Frac{( \, u_{\alpha} \, - \,
    \phi_{\alpha}) \, x}{R\,v}
\end{equation}
where $u_{\alpha} \equiv R_{\alpha}/R$ is the number fraction of
$\alpha$ grains in the active layer. Now (\ref{final3}), and
(\ref{final4}) can be used to express $\phi_{\alpha}(x)$ in terms of
$u_{\alpha}$ and obtain the solution of (\ref{pred7}). The constant of
integration for this solution must be evaluated by enforcing the loading
condition given by (\ref{load}).

        We consider  various cases where the grains of the two species
differ either in size or in surface properties. The mean number fraction
$\Phi_{\alpha}$ has been set to $0.5$ in all the cases.

\subsection{Grains differing only in surface properties} \label{pred:surf}

        It is reasonable to suppose that amplification is largely guided
by differences in sizes while capture is driven by differences in
surface properties. Then for this case, we may set
\begin{equation*}
  \gamma^{a}_{\alpha \alpha} = \gamma^{a}_{\alpha \beta} =
  \gamma^{a}_{\beta \alpha} = \gamma^{a}_{\beta \beta} = \gamma^{a}.
\end{equation*}
        The differences in surface properties would, of course, lead to
the capture coefficients being different, and we set
\begin{equation*}
  \gamma^{c}_{\beta}\, = \, \gamma^{c}, \: \gamma^{c}_{\alpha} \, = \,
  s \, \gamma^{c}_{\beta}.
\end{equation*}
A value of $s \, > \, 1$ implies that $\alpha$ type grains are rougher
than $\beta$, and hence more easily captured by the passive layer. With
these simplifications, it follows that the solution of
(\ref{final2})--(\ref{final4}) is:
\begin{equation}
\label{pred8}
  \phi_{\alpha}\:=\: \Frac{s \, u_{\alpha}}{1 \: + \: (s-1) \, u_{\alpha}}
\:\: \mbox{and} \:\:
  \Frac{u_{\alpha}}{(1\: - \: u_{\alpha})^{s}} \: = \: C_{1}\, (1\, -\,
  x^2)^{s-1}
\end{equation}
where $C_{1}$, the constant of integration is obtained by numerical
integration of (\ref{load}). Figures 3 and 4 present the variation of the
number fraction of $\alpha$ type grains in the active and the passive
layers, respectively, along the length of the interface for two cases, one
where the $\alpha$ type grains are slightly rougher than the $\beta$ type
grains ($s\,=\,1.5$), and another where the $\alpha$ type grains are much
rougher ($s\,=\,10$). The model predicts depletion of the rougher grains at
the two ends of the drum and accumulation near the center of the drum. The
degree of segregation, of course, depends upon the difference between the
surface properties of the two species, quantified in this model by $s$.
\begin{figure}
\xlabel{$x$}
\ylabel{$u_{\alpha}$}
\includegraphics[width=8cm]{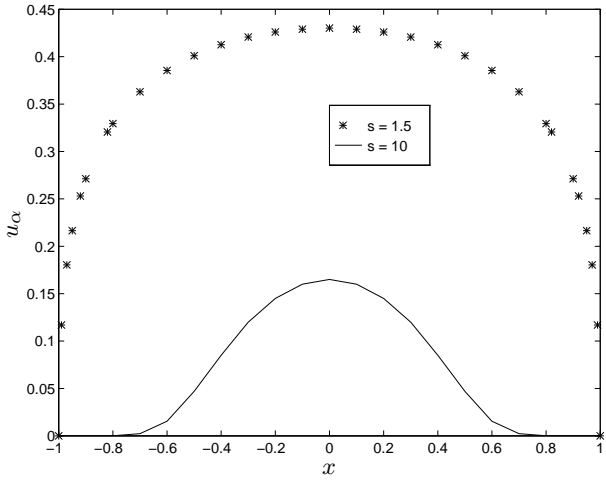}
\vspace{0cm}
\caption{Variation of the number fraction of $\alpha$ type grains (the
rougher species) in the active layer along the length of the interface
for the case of grains differing only in surface properties. The
relative roughness of the $\alpha$ species increases with the parameter
$s$.}
\label{shap_r1}
\end{figure}
\begin{figure}
\xlabel{$x$}
\ylabel{$\phi_{\alpha}$}
\includegraphics[width=8cm]{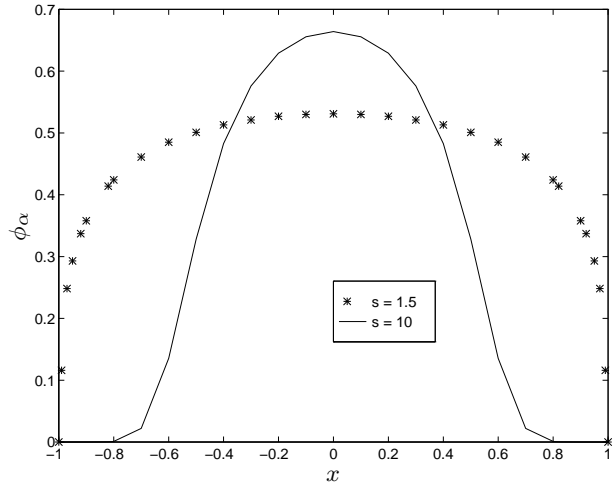}
\vspace{0cm}
\caption{Variation of the number fraction of $\alpha$ type grains in
the passive layer along the length of the interface, for the same
parameter set as in Fig.\ 3.}
\label{shap_fi1}
\end{figure}

\subsection{Grains differing slightly in size} \label{pred:size_small}.
        Following  our  arguments  in \S$\,$\ref{pred:surf},  we  will
assume that the difference in size leads to a difference in the rate of
amplification of the two species but not that of capture. We investigate
the case where grains have identical surface properties but differ slightly
in size. For the case of grains differing neither in size nor surface
properties (\emph{i.e.} identical species), the solution is, of course,
that of constant number fraction in the active and passive layers,
\begin{equation}
  \label{pred10} u_{\alpha}(x) \; = \; \phi_{\alpha}(x) \; = \;
  \Phi_{\alpha},
\end{equation}
where $\Phi_{\alpha}$ is the mean number fraction of $\alpha$ type grains
during filling of the drum. For a mixture comprising species that differ
very slightly in size, we allow the amplification constants for the two
species to differ by a small amount $\vartheta$. The probability of a large
rolling grain dislodging a large immobile grain is equal to that of a small
rolling grain dislodging a small immobile grain. Hence,
\begin{equation}
  \label{pred12} \gamma^{a}_{\alpha \alpha} \, = \, \gamma^{a}_{\beta
  \beta} \, = \, \gamma^{a}.
\end{equation}
However, if the $\beta$ type grains are larger, the probability of a
$\beta$ type grain dislodging an $\alpha$ type grain is greater than that
of it dislodging one of its kind, and the probability that an $\alpha$ type
grain dislodges a $\beta$ type grain is less than that of it dislodging one
of its kind. Hence, for small differences in sizes, we set
\begin{equation}
  \label{pred13} \gamma^{a}_{\alpha \beta} \, =  \, \gamma^{a} \, (1
  \, + \, \vartheta),
\end{equation}
\begin{equation} \label{pred14}
  \gamma^{a}_{\beta \alpha} \, =  \, \gamma^{a} \, (1 \, - \,
  \vartheta).
\end{equation}
        We now seek the solution as a perturbation to the uniform
solution (\ref{pred10}) to leading order in $\vartheta$,
\begin{equation}
  \label{pred15} u_{\alpha}(x) \, =  \, \Phi_{\alpha} \, + \,
  \vartheta \, u'_{\alpha}(x), \:\: \phi_{\alpha}(x) \, =  \,
  \Phi_{\alpha} \, + \, \vartheta \, \phi'_{\alpha}(x)
\end{equation}
and solve for $u'_{\alpha}(x)$ and $\phi'_{\alpha}(x)$. Substituting
(\ref{pred12}) -- (\ref{pred15}) in (\ref{final3}) -- (\ref{final4}), we
can express $\phi'_{\alpha}(x)$ in terms of $u'_{\alpha}(x)$ and $x$ as
\begin{equation}
  \label{pert1} \phi'_{\alpha}(x) \, = \, u'_{\alpha}(x) \, - \,
\gamma^{c} \, \Phi_{\alpha} \, \Phi_{\beta} \, (\Frac{\gamma^{c} \, Rv
\, - \, x}{\gamma^{c \, 2} \, Rv \, + \, \gamma^{a} \, x}) \,
\Frac{x}{Rv}.
\end{equation}
Using (\ref{pred15}), we can write (\ref{pred7}) as:
\begin{equation}
  \label{pert2}    \der{u'_{\alpha}}{x} \: = \: \Frac{( \, u'_{\alpha}
  \, - \, \phi'_{\alpha} \,) \, x}{R\,v}.
\end{equation}

        We now consider special cases for which simple analytical
solutions of (\ref{pert2}) are possible.  The first is the case of the
grains of both species being very smooth, \emph{i.e}
$\textfrac{\gamma^{c}}{\gamma^{a}} \, \ll \, 1$. The solution is
\begin{eqnarray} \label{pert6}
u_{\alpha}(x)  & = &  \Phi_{\alpha} \, + \vartheta \,
\Frac{\gamma^{c}}{\gamma^{a}} \, \Phi_{\alpha} \, \Phi_{\beta}
\ln(1-x^2),\\ \phi_{\alpha}(x)  & = &   \Phi_{\alpha} \, + \,
\vartheta \, \Frac{\gamma^{c}}{\gamma^{a}} \, \Phi_{\alpha} \,
\Phi_{\beta} \, [1\,+\,\ln(1-x^2)].
\end{eqnarray}
Figures 5 and 6 present these results for $\vartheta = 0.1$, where it is
clear that the central core is richer in smaller grains, in agreement
with experimental observations.
\begin{figure}
\xlabel{$x$} \ylabel{$u_{\alpha}\, , \, u_{\beta}$}
\includegraphics[width=8cm]{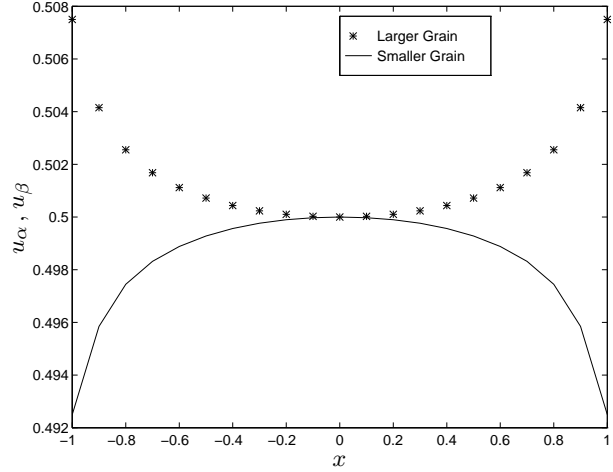}
\vspace{0cm}
\caption{Variation of the number fraction of $\alpha$ (smaller) and
$\beta$ (larger) type grains in the active layer along the interface.
For this case, both species are very smooth, and the parameter
$\vartheta$ is set to 0.1.}
\label{pert1_u}
\end{figure}
\begin{figure}
\xlabel{$x$} \ylabel{$\phi_{\alpha} \, , \, \phi_{\beta}$}
\includegraphics[width=8cm]{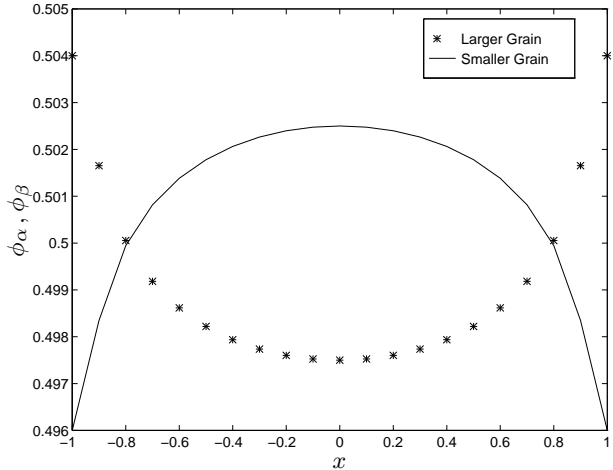}
\vspace{0cm}
\caption{Number fraction of $\alpha$ (smaller) and $\beta$ (larger)
type grains in the passive layer as a function of position on the
interface. Parameter values the same as in Fig.\ 5.}
\label{pert1_fi}
\end{figure}

        The second case is when both species are very rough, \emph{i.e}
$\textfrac{\gamma^{c}}{\gamma^{a}} \, \gg \, 1$. The solution is
\begin{eqnarray} \label{pert11}
u_{\alpha}(x) & = & \Phi_{\alpha} \, - \vartheta \, \Phi_{\alpha}
\,\Phi_{\beta} \ln(1-x^2), \\ \phi_{\alpha}(x) & =  & \Phi_{\alpha} \,
- \, \vartheta \, \Phi_{\alpha} \, \Phi_{\beta} \, [1\,+\,\ln(1-x^2)].
\end{eqnarray}
Figures 7 and 8 present the profiles of $u_{\alpha}$ and $\phi_{\alpha}$
for $\vartheta = 0.1$, where it is apparent that, unlike in the case of
smooth grains, the smaller grains are depleted near center of the drum
and tend to accumulate at the two ends. This is due to the fact that if
both species are equally rough, the larger grains will roll further down
the active layer as their inertia is greater. No experimental results
are however available to validate this prediction.
\begin{figure}
\xlabel{$x$} \ylabel{$u_{\alpha}\, , \, u_{\beta}$}
\includegraphics[width=8cm]{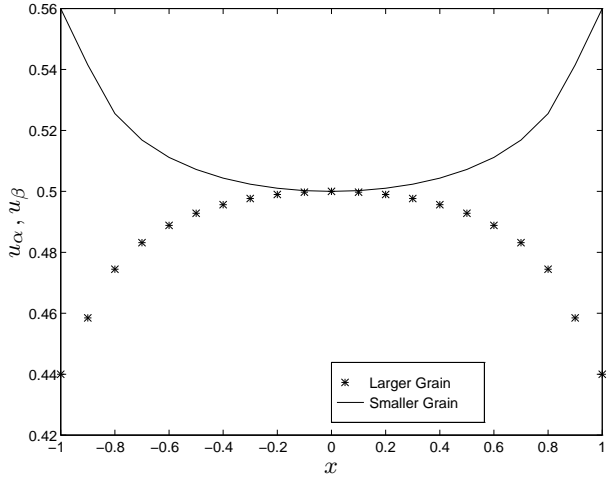}
\vspace{0cm}
\caption{Profiles of number fractions of $\alpha$
(smaller) and $\beta$ (larger) type grains in the active layer when both
species are very rough, for $\vartheta=0.1$}
\label{pert2_u}
\end{figure}
\begin{figure}
\xlabel{$x$} \ylabel{$\phi_{\alpha} \, , \, \phi_{\beta}$}
\includegraphics[width=8cm]{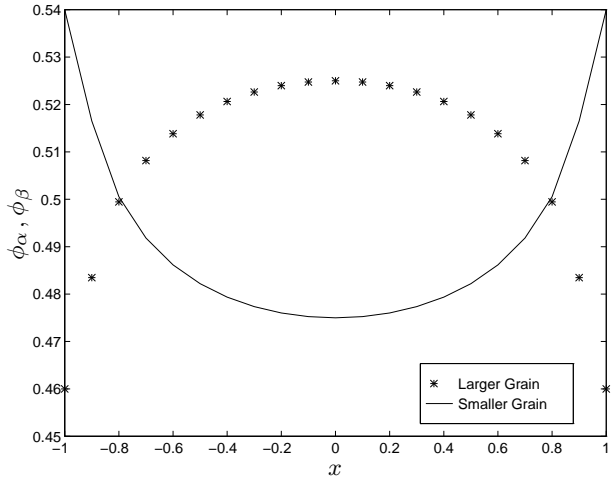}
\vspace{0cm}
\caption{Profiles of number fraction of $\alpha$
(smaller) and $\beta$ (larger) type grains in the passive layer at the
interface, for the parameter set as in Fig. \ 7.}
\label{pert2_fi}
\end{figure}

\subsection{Grains differing widely in size} \label{pred:perc}
        For large differences in sizes between the two species,
experimental observations in sandpiles reveal either complete
segregation or stratification; it has been reported \cite{makse_etal}
that this occurs when the size ratio exceeds 1.5. Boutreux
\etal\ \cite{boutreux_etal}proposed that size segregation is then due to
percolation or ``kinematic sieving'', as the smaller rolling grains tend
to fall through the gaps between the large grains, forming a sub-layer
above the interface. The larger rolling grains are therefore not in
contact with the bulk, and are not captured to the extent they would
have been if in contact. To account for this phenomenon, we follow
\cite{boutreux_etal} and modify $\Gamma_{\alpha}$ by a factor $\exp(p \,
u_{\alpha})$, which mimics the screening of the large grains from the
interface by the sub-layer of the small grains. The dimensionless
parameter $p$, which is proportional to the size ratio of $\beta$ to
$\alpha$ type grains, measures the degree of percolation.

        For a large difference in sizes, segregation due to percolation effect
would  far  exceed  the  segregation  effects due  to  differences  in
amplification    and   capture    coefficients,   as    suggested   by
\cite{boutreux_etal}. Hence we assume,
\begin{equation} \label{perc1}
  \gamma^{a}_{\alpha \alpha} = \gamma^{a}_{\alpha \beta} =
  \gamma^{a}_{\beta \alpha} = \gamma^{a}_{\beta \beta} = \gamma^{a}.
\end{equation}
and
\begin{equation} \label{perc2}
  \gamma^{c}_{\alpha}\, = \, \gamma^{c}_{\beta} \: = \; \gamma^{c}.
\end{equation}
Using (\ref{perc1}) and (\ref{perc2}), we solve (\ref{final3}) --
(\ref{final4}) to obtain:
\begin{equation}
\label{perc3}
\phi_{\alpha}(x) \, = \, \Frac{u_{\alpha}(x)}{u_{\alpha}(x) \, + \, (1 \, - \,
  u_{\alpha}(x)) \, \exp[p \, u_{\alpha}(x)]}.
\end{equation}
and solve (\ref{pred7}) numerically. The results shown in figures 9 and
10 for $p\,=\,2$ indicate that the larger grains are heavily depleted in
the central region of the drum and accumulate at the ends both in the
active and passive layers. This is in qualitative agreement with
experimental observations \cite{nakagawa_etal}.
\begin{figure}
\xlabel{$x$}
\ylabel{$u_{\alpha}\, , \, u_{\beta}$}
\includegraphics[width=8cm]{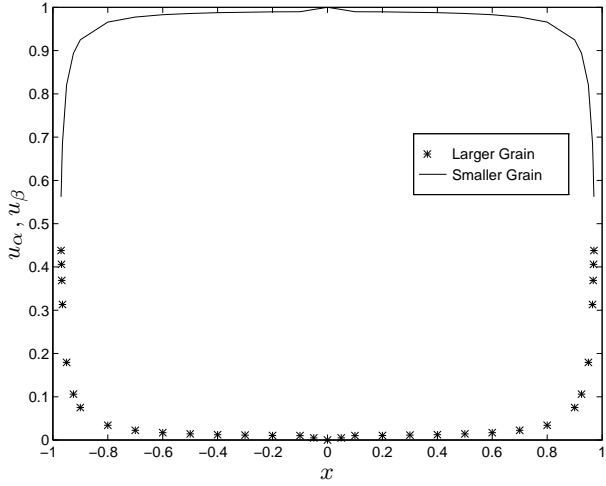}
\vspace{0cm}
\caption{Profiles of the number fraction of $\alpha$ (smaller) and
$\beta$ (larger) type grains in the active layer for the case of grains
differing widely in size. See text for parameter values}
\label{perc_u}
\end{figure}
\begin{figure}
\xlabel{$x$}
\ylabel{$\phi_{\alpha} \, , \, \phi_{\beta}$}
\includegraphics[width=8cm]{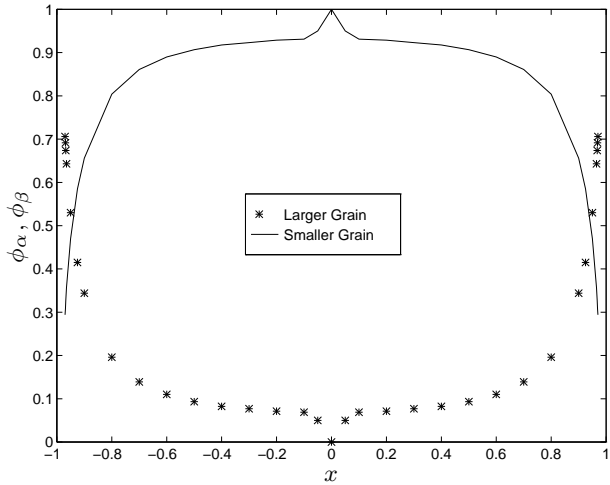}
\vspace{0cm}
\caption{Profiles of the number fraction of $\alpha$ (smaller) and
$\beta$ (larger) type grains in the passive layer at the interface for
the case as in Fig.\ 9.}
\label{perc_fii}
\end{figure}

\subsection{Other cases: Model inconsistency} \label{symmetry}
The solutions described in the previous section are symmetric about the
mid-plane $x=0$. However this is not a generic result, rather only for the
special cases we have considered, and also because the slope functions have
been chosen to satisfy $g^a + g^c = \mbox{constant}$. If the grains differ
in size \emph{and} shape, or if the slope functions do not satisfy the
above constraint, the solutions for $R_i$ and $\phi_i$ are not symmetric.
This result is inconsistent with our assumption that the passive layer
executes solid body rotation: circular streamlines in the passive layer
necessitates symmetry of $R_i$ and $\alpha_i$

        It thus appears that the assumption of solid body rotation in the
passive layer must be relaxed for consistency, and solutions in general
need not be symmetric about the mid-plane. This may be accomplished by
postulating an interface of finite thickness within which streamlines
are not circular, which is the approach we are following currently and
hope to report the results soon. The recent study of Shinbrot \etal\
\cite{shinbrot_etal} reports the existence of an interface between the
active and the passive layers, wherein deformation occurs by periodic
stick-slip motion. In this light, further experimentation is necessary
to ascertain whether the concentration of species in the active and
passive layers are symmetric.

\section{Conclusions \label{c:conc}}
        Granular mixtures, consisting of grains that differ in size and/or
surface properties, have been observed to undergo segregation when they
are either poured in a heap or rotated in a drum. Several analytical
models have been proposed to predict segregation in case of a two
dimensional sand-pile. The theoretical formalism proposed by Bouchaud
\etal\ \cite{bcre} and later extended by
\cite{boutreux_degennes,makse97,boutreux_etal}, successfully addresses
the problem of segregation in poured heaps. The present work attempts to
extend this framework to the rotating drum problem.

        We argue that the qualitative nature of segregation of two mixtures,
one of which differ in surface properties and the other in size, may be
quite different even if their angles of repose are equal. Our model
recognizes collisional interactions between grains at the interface as
the cause for segregation. Such interactions are significantly
influenced by the size of grains and their surface properties, apart
from the angle of repose. This provides a canonical framework under
which segregation of grains differing in size \emph{and} shape could be
studied in various geometries. The model gives satisfactory predictions
for grains differing either in surface properties or size. These results
are in good qualitative agreement with experimental observations. For
the case of segregation due to small differences in size of grains when
both the species are very rough, the prediction of our model is
counter-intuitive
- such a system has not yet been studied experimentally, and this work
provides a motivation for it.

        Our predictions of symmetric profiles of the species concentrations
in the active and passive layers are only for particular cases of grain
properties. This is not always the case, and the solution is in general
not symmetric; this is an inconsistency in the model as it is contrary
to the assumption of circular streamlines in the passive layer. We
believe that the inconsistency arises from the assumption of a sharp
interface between the active and passive layers, and are currently
working to incorporate a finite sized interface in which streamlines
deviate from that of rigid body motion. This will result in asymmetric
profiles of the species concentration. While recent experiments suggest
asymmetry in the active layer thickness, more experimental studies are
necessary to resolve this issue.


\begin{thebibliography}{10}

\bibitem{metcalfe_etal}
G. Metcalfe, T. Shinbrot, J. J. McCarthy and J. M. Ottino, {\it
Nature} {\bf 374}, (1995) 39.

\bibitem{henien_etal}
H. Henien, J. K. Brimacombe and A. P. Watkinson,
{\it Metall. Trans. B} {\bf 16B}, (1985) 763.

\bibitem{nityanand_etal}
M. Nityanand, B. Manley and H. Henien,
{\it Metall. Trans. B} {\bf 17B} (1983) 247.

\bibitem{clement_etal}
E. Cl\'{e}ment, J. Rajchenbach and J. Darun,
{\it Europhys.\ Let.} {\bf 30} (1995) 7.

\bibitem{cantelaube_bideau}
F. Cantelaube and D. Bideau,
{\it Europhys.\ Let.} {\bf 30} (1995) 133.

\bibitem{dasgupta_etal}
S. Dasgupta, D. V. Khakhar and S. K. Bhatia,
{\it Chem. Engng. Sci.} {\bf 46} 576, (1991) 1513.

\bibitem{nakagawa_etal}
M. Nakagawa, S. A. Altobelli, A. Caprihan and E. Fukushima,
{\it Chem. Engng Sci.} {\bf 52} 23, (1997) 4423.

\bibitem{khakhar_etal}
D. V. Khakhar, J. J. McCarthy and J. M. Ottino,
{\it Physics of Fluids} {\bf 9} 12, (1997) 3600.

\bibitem{boateng_barr96}
A. A. Boateng and P. V. Barr,
{\it Chem.\ Engng Sci.} {\bf 51} 17, (1996) 4167.

\bibitem{boateng_barr97}
A. A. Boateng and P. V. Barr, {\it J.\  Fluid Mech.} {\bf 330},
(1997) 233.

\bibitem{bcre}
J. P. Bouchaud, M. E. Cates, J. Ravi Prakash and
S. F. Edwards {\it J. Phys.\ I France} {\bf 4}, (1994) 1383.

\bibitem{boutreux_degennes}
T. Boutreux and P. G. de Gennes, {\it J. Phys.\ I France} {\bf 4},
(1996) 1295.

\bibitem{makse97}
H. A. Makse,
{\it Phys. Rev. {E}} {\bf 56} 6, (1997) 7008.

\bibitem{boutreux_etal}
T. Boutreux, H. Makse and P. G. de Gennes,
Submitted to {\it Eur.\ Phys. J. B}, (1998).

\bibitem{makse98}
H. A. Makse, {\it Cond-mat 9809422}, (1998).

\bibitem{johnson_etal}
P. C. Johnson, P. Nott and R. Jackson,
{\it J. Fluid Mech.} {\bf 210}, (1990) 501.

\bibitem{makse_etal}
H. A. Makse, S. Havlin, P. R. King and H. E. Stanley, {\it Nature},
{\bf 386}, (1997) 379.

\bibitem{shinbrot_etal}
T. Shinbrot, A. Alexander and F. J. Muzzio,
{\it Nature} {\bf 397}, (1999) 675.
\end{thebibliography}

\end{document}